\documentclass[pageno]{jpaper}

\usepackage[normalem]{ulem}
\usepackage{algorithm}  
\usepackage{algorithmicx} 
\usepackage{algpseudocode} 
\usepackage{threeparttable}
    
\usepackage{url}

\begin{document}

\title{HMM-V: Heterogeneous Memory Management for Virtualization}

\author{Sai sha \\Peking University\vspace{-2cm}
\and Chuandong Li\\Peking University\vspace{-2cm}\\
\and Yingwei Luo\\Peking University\vspace{-2cm}\\
\and Xiaolin Wang\\Peking University\vspace{-2cm}\\
\and  \\
\and Zhenlin Wang\\Michigan Technological University\vspace{-2cm}\\
}
\date{}
\maketitle
\thispagestyle{empty}

\begin{abstract}

The memory demand of virtual machines (VMs) is increasing, while DRAM has limited capacity and high power consumption. Non-volatile memory (NVM) is an alternative to DRAM, but it has high latency and low bandwidth. We observe that the VM with heterogeneous memory may incur up to a $1.5\times$ slowdown compared to a DRAM VM, if not managed well. However, none of the state-of-the-art heterogeneous memory management designs are customized for virtualization on a real system.

In this paper, we propose HMM-V, a Heterogeneous Memory Management system for Virtualization. HMM-V automatically determines page hotness and migrates pages between DRAM and NVM to achieve performance close to the DRAM system. First, HMM-V tracks memory accesses through page table manipulation, but reduces the cost by leveraging Intel page-modification logging (PML) and a multi-level queue. Second, HMM-V quantifies the ``temperature'' of page and determines the hot set with bucket-sorting.
HMM-V then efficiently migrates pages with minimal access pause and handles dirty pages with the assistance of PML. Finally, HMM-V provides pooling management to balance precious DRAM across multiple VMs to maximize utilization and overall performance. 
HMM-V is implemented on a real system with Intel Optane DC persistent memory. The four-VM co-running results show that HMM-V outperforms NUMA balancing and hardware management (Intel Optane memory mode) by $51\%$ and $31\%$, respectively.

\end{abstract}

\section{Introduction}

%大内存需求，DRAM功耗、成本也高
Traditional DRAM has limited capacity and expensive price per unit capacity, as well as high power consumption because of real-time flushing. In fact, DRAM consumes about $40\%$ of power in modern data centers~\cite{2009PDRAM,2003Energy,2009Scalable}.
%NVM的优势和缺点
Large-capacity non-volatile memory (NVM) is an effective alternative to DRAM. NVM has no flush and idle power consumption and supports byte-addressable access~\cite{2018Big}. In particular, the Intel Optane DC persistent memory (PMem) is commercially available. However, compared to DRAM, NVM comes with lower bandwidth and longer latency~\cite{raybuck2021hemem,yao2020matrixkv,dong2019performance,basic-test}. 

%问题
The fast DRAM and slow NVM make up a heterogeneous memory system. The researchers focus on putting active (hot) pages in DRAM and inactive (cold) pages in NVM for the best performance. The classic management approach achieves the goal by page tracking, classifying, and migration.
Following this design, many advanced techniques have been developed, such as {\it HeMem}~\cite{raybuck2021hemem}, {\it Nimble}~\cite{yan2019nimble}, {\it HeteroOS}~\cite{HeteroOS}, {\it RAMinate}~\cite{hirofuchi2016raminate}, etc. Unfortunately, all of them cannot provide efficient management for virtualization on a real heterogeneous memory system. 
Heterogeneous memory management suffers from new challenges in virtualization.
First, multi-VM co-running can generate intensive memory competition while performance isolation between VMs is essential. For example, by using total DRAM as a direct-mapped cache, the Intel Optane's memory mode ({\it MM})~\cite{OptaneDC}  can hide the latency of PMem access (§\ref{pmem}). However, the experiments show that multi-VM co-running can increase DRAM cache misses by $2\times$ to $4\times$ than a stand-alone VM, because of severe DRAM cache pollution (§\ref{subex:multivmwiththp}).
Second, the virtualization introduces VM context, and we should minimize expensive VMTraps (i.e., context switch between guest and host). For example, the write protection-based page migration adopted by {\it HeMem}, incurs heavy VMTraps due to write exception handling. Finally, VM memory overcommit~\cite{gordon2011ginkgo} in a heterogeneous memory system suffers from a new challenge comparing to traditional single-DRAM system. The large-capacity NVM guarantees sufficient VM memory capacity, but the fast DRAM is scarce. Full use of DRAM is the key to maintain performance. Ideal management should dynamically balance DRAM across multiple VMs for better overall performance.

%Virtualization introduces new opportunities for heterogeneous memory management. First, memory virtualization introduces two-dimensional (2D) address mapping. By modifying the guest physical address to host physical address~\cite{ssp}, we can flexibly and transparently adjust the memory of VM applications. Then the hardware-assisted virtualization technologies can be utilized for optimization. One example is Intel page-modification logging (PML)~\cite{pml-inteldoc}, which is designed for tracking dirty pages for VM live migration~\cite{bitchebe2021extending}. Leveraging PML, we can efficiently handle dirty pages with few VMTraps during page migration.
%本段需要细细琢磨，重写
Virtualization introduces new opportunities for heterogeneous memory management. (1) We leverage hardware-assisted virtualization technologies to optimize our designs. Intel page-modification logging (PML) is designed for tracking dirty pages for VM live migration~\cite{pml-inteldoc,bitchebe2021extending}. By using PML, we can trace the accessed (dirty) page tables rather than the entire page tables to improve memory access tracking efficiency. Moreover, PML also can efficiently handle dirty pages with few VMTraps when migrating pages between DRAM and NVM.
(2) Memory virtualization introduces two-dimensional (2D) address mapping. By manipulating the mappings of guest physical addresses to host physical addresses, we can transparently resize VM DRAM according to the memory access patterns. Particularly, we can utilize page migration and remapping rather than the inefficient ballooning to achieve the multi-VM DRAM balancing.

\begin{figure}[t]
\begin{center}
\includegraphics[width=0.5\textwidth]{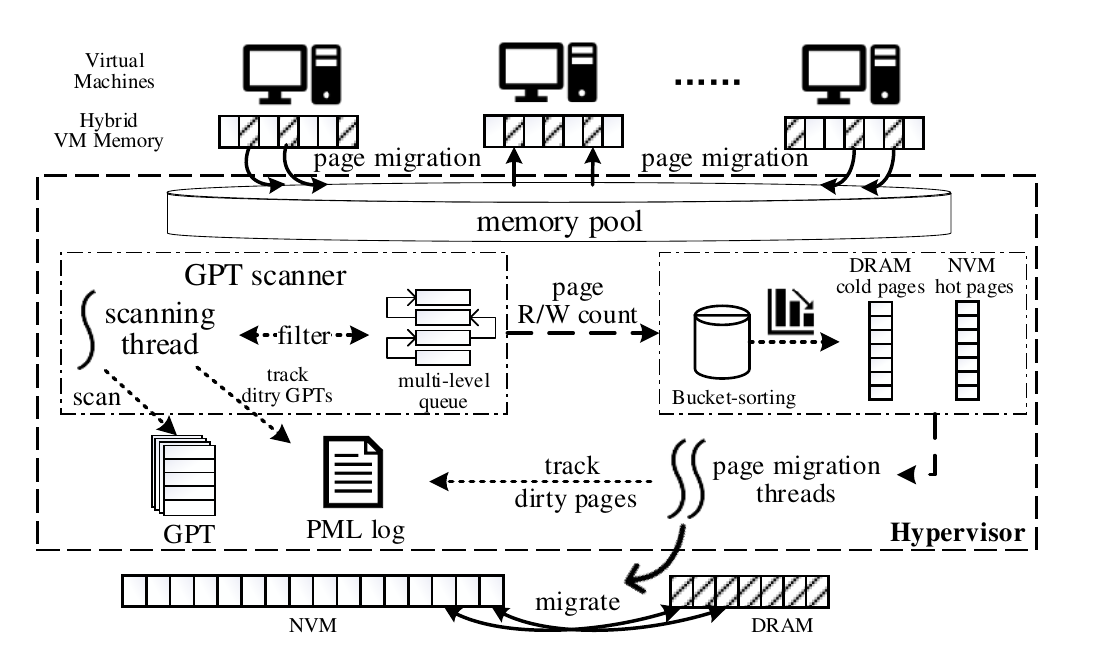}
\caption{HMM-V system design.}   
\label{fig:arch}
\end{center}
\vspace{-0.5cm}
\end{figure}

%our design
In this paper, we present {\it HMM-V}, a heterogeneous memory management system for virtualization. Figure~\ref{fig:arch} shows the high-level overview of HMM-V. HMM-V manages heterogeneous memory in the hypervisor and follows the classic approach (i.e., page tracking, classifying, and migration) to manage heterogeneous memory, but with novel optimization in every step.
%page tracking
First, HMM-V tracks VM page accesses by periodically scanning guest pages tables (GPTs) to record and clear accessed/dirty (A/D) bits. HMM-V only scans dirty GPTs logged by PML instead of the entire GPTs, which effectively reduces the page table scan time. Scanning page table A/D bits has overhead because of setting bits and TLB shootdowns for pages. HMM-V adopts a multi-level queue, where more frequently accessed pages are placed in higher levels and tracked less frequently. This locality-based design can reduce tracking cost while preserving page access feature.

%hot or cold
HMM-V screens out the hottest pages and flexibly determines the hot set according to the VM DRAM size by sorting pages. HMM-V adopts a fine-grain approach which measures the {\it page-degree} (i.e., the ``temperature'' of a page) and bucket-sorts VM pages based on {\it page-degree}s. The{\it page-degree} of a page is a weighted combination of its read and write frequency where the weights are derived from read and write costs, respectively. 

%page migration
HMM-V migrates pages by utilizing PML again. It copies data in parallel with minimal access pause. Then, handles dirty pages by checking the PML log. Compared to write protection-based page migration adopted by {\it HeMem}, our design generates few VMTraps. In addition, HMM-V actively fills the new mappings of Intel extended page table (EPT) during migration to avert VMTraps caused by the EPT page faults~\cite{pml-inteldoc,epttest}. 

%pool
Finally, HMM-V manages co-running VMs and balances DRAM by pooling. The pool extracts the surplus DRAM from the VMs whose DRAM sizes are bigger than their hot set sizes, as well as, gives it to the VMs with insufficient DRAM. HMM-V thus improves memory utilization and overall performance. HMM-V utilizes page migration instead of ballooning to efficiently and transparently adjust VM memory for supporting memory overcommit in heterogeneous memory systems.

We implement and evaluate HMM-V in an Intel Xeon(R) Gold server with PMem. In addition, we extend HMM-V to support transparent huge page (THP) management~\cite{hpvmm}.
The multi-VM experimental results show that in the regular page system, compared with NUMA balancing~\cite{gaud2014large}, which indicates that NUMA balancing is enabled in VMs without other management, HMM-V provides $51\%$ higher performance, meanwhile, HMM-V outperforms Intel {\it MM} by $31\%$. For the THP system, HMM-V achieves $143\%$ higher performance than NUMA balancing and $14\%$ higher performance than Intel {\it MM} and $615\%$ higher performance than {\it Nimble Page Management}~\cite{yan2019nimble}.

In summary, the contributions of our work are:
\begin{itemize} 
\item We propose a PML-based GPT scanner to track the pages of the active processes only and design a multi-level queue to reduce page tracking overhead.
\item We propose a hot/cold page classifier based on sorting, which can adapt to various memory access patterns. 
\item We migrate pages in parallel with minimal access pause and efficiently handle dirty pages by leveraging PML.
\item We propose memory pooling management to effectively balance DRAM among VMs for improving utilization and overall performance when multi VMs co-run.
\item We implement HMM-V on a real system and extend HMM-V to support THP.
\item We extensively evaluate HMM-V and compare it with several state-of-the-art designs.
\end{itemize}

\begin{table*}[ht]
%\vspace{-0.5cm}
 \centering
  \caption{Comparison of heterogeneous memory management}
   \label{tab:relatework}%
%  \begin{threeparttable}
    \small
    \begin{tabular}{rccccc}
    \toprule
          & \textbf{RAMinate} & \textbf{HeteroOS} & \textbf{Nimble} & \textbf{HeMem} & \textbf{HMM-V} \\
    \hline
    VM support & yes   & yes   & no    & no    & yes \\
    %unmodify gOS & yes   & no    & -     & -     & yes \\
    memory track & EPT scan & GPT scan & PT scan & PEBS  & GPT scan (optimized) \\
    hotset policies & LRU-based queue & LRU of Linux  & LRU of Linux  & fixed threshold & sorting (adaptive) \\
    \multirow{2}[0]{*}{page migration} & \multicolumn{2}{c}{low speed} & high speed but & high speed but & high speed and \\
          & \multicolumn{2}{c}{(raw Linux migration)} & long access pause & heavy VMTraps  & few VMTraps \\
    DRAM balancing & no    & yes  & no    & no    & yes \\
     real system   & no    & no    & no    & yes   & yes \\
    \bottomrule
    \end{tabular}%
 %   \begin{tablenotes}
  %      \footnotesize
 %       \item[*] inefficient ballooning. ** pooling by leveraging  efficient page migration.
 % \end{tablenotes}
 %   \end{threeparttable}
\end{table*}%

\section{Related Work}
%-------------------------------------------------------------
\label{sec:relatedwork}
Table~\ref{tab:relatework} compares HMM-V with several state-of-art designs of heterogeneous memory management.

{\it RAMinate}~\cite{hirofuchi2016raminate} is a heterogeneous memory management system for hypervisor-based virtualization. It forces on reducing write traffic to NVM by optimizing page locations. {\it RAMinate} tracks memory accesses by periodically scanning the entire EPT, logging and clearing A/D bits (called the EPT scanner). Then, {\it RAMinate} flushes TLB to ensure that each access triggers a page walk and sets the A/D bits. The results show that, without optimization, the EPT scanner incurs up to a $20\%$ slowdown (§\ref{sub:ex-pagetrack}). 

{\it HeteroOS}~\cite{HeteroOS} is an OS-level solution for virtualization. For optimization, {\it HeteroOS} sacrifices the principle of full virtualization and exposes heterogeneous features to VMs. In addition, {\it HeteroOS} extends ballooning to support memory overcommit. However, the ballooning in a heterogeneous memory architecture is no longer appropriate because it is slow, especially for adjusting regular page memory~\cite{HUB}, and gOS needs to be customized to distinguish DRAM and NVM.

{\it Nimble}~\cite{yan2019nimble} is a tiered memory management solution that supports the THP mechanism in non-virtualization. {\it Nimble} achieves THP migration without page demotion. Although {\it Nimble} provides high migration throughput through parallel, it incurs long page access pause during migration.
Both {\it HeteroOS} and {\it Nimble} use Linux active/inactive LRU list~\cite{bovet2005understanding} to track and distinguish hot/cold pages. However, this design is not sufficient to track diversified memory access patterns. We will show this through experiments in §\ref{sub:ex-thp}.

{\it HeMem}~\cite{raybuck2021hemem} is a tiered main memory management system designed for big data applications. It is designed as a user-level library and handles page faults using userfaultfd~\cite{userfaultfd}, which makes it not apply to virtualization. {\it HeMem} uses PEBS~\cite{pebs-inteldoc} to track memory access and determines the hot set with a fixed threshold. We observe that the fixed threshold cannot adapt the diverse memory access patterns (§\ref{sub:ex-ablationstudy}). For migration, {\it HeMem} uses page write-protection to reduce page access pause. However, in virtualization, write-protection triggers expensive VMTraps, for example, shadow paging synchronization~\cite{ssp}. Except for {\it HeMem}, other designs are evaluated in a simulated heterogeneous memory system.

In contrast, {\it HMM-V} is customized for virtualization and achieves efficient heterogeneous memory management in a real system. In particular, {\it HMM-V} utilizes virtualization features to optimize page manipulations and reduce management overhead.

\section{Background}
\subsection{Intel Optane DC Persistent Memory}
\label{pmem}
%The NVM emerged in recent years, such as phase change memory (PCM)~\cite{lee2009architecting}, spin-transfer torque magnetic random access memory (STT-MRAM)~\cite{apalkov2013spin}, etc. %With huge capacity, random access, high storage density, strong scalability and no flushing power consumption, NVM has become an alternative to DRAM.
Intel Optane DC PMem~\cite{OptaneDC} is the first and only commercially-available NVM product. There are two configurations for PMem: memory mode ({\it MM}) and APP Direct mode~\cite{pmdoc}. {\it MM} presents a typical hierarchical architecture and all the DRAM of a socket is configured as a direct-mapped cache. The DRAM cache is transparent to users and {\it MM} manages DRAM and PMem with hardware. In contrast, in APP Direct mode, DRAM and PMem are at the same tier, which is a horizontal architecture. The PMem can be mounted as a separate NUMA node along with DRAM as main memory, and our design is based on this configuration. However, PMem comes at cost of up to $6\times$ lower bandwidth and up to twice the latency compared with DRAM. Table~\ref{tab:dramvs.nvm} shows the comparison in detail. The write latency of PMem is close to DRAM, but PMem's write bandwidth is only about one-sixth of DRAM. Moreover, NVM latency and bandwidth show significant asymmetry for read and write.  Based on this observation, we assign different read and write weights when distinguishing between hot and cold pages (§\ref{subsec:pagedegree}).
 
\begin{table}[h!]
  \vspace{-0.5cm}
  \centering
  \caption{DRAM and PMem latency and bandwidth.}
    \begin{tabular}{ccc}
    \toprule
    \textbf{Memory} & \textbf{R/W Latency(ns)} & \textbf{R/W BW(GB/s)} \\
    \midrule
    DRAM  & 81 / 82 & 120 / 82 \\
    PMem  & 310 / 94 & 37 / 13 \\
    \bottomrule
    \end{tabular}%
  \label{tab:dramvs.nvm}%\
  \vspace{-0.5cm}
\end{table}%

\subsection{Intel Page-Modification Logging (PML)}
%28.3.6 Page-Modification Logging
\label{subsec:pml}
%PML的起源和作用
\vspace{-0.2cm}
Intel PML is designed to optimize dirty page handling in VM live migration~\cite{pml,bitchebe2021extending}. During live migration, VM memory accesses are not interrupted. Dirty pages during migration need to be tracked and re-transmitted. To track dirty pages, in the past, the hypervisor write-protects all VM pages, thus writing triggers to write exception. An expensive VMTrap is triggered to handle this exception in the hypervisor and a dirty page bitmap is updated. 

%PML的使用方法
PML tracks dirty pages of a VM in hardware without triggering expensive VMTraps. Before a guest-physical access, the processor may determine that it first needs to set an accessed or dirty bit for its EPT entry (EPTE)~\cite{pml-inteldoc}. If Intel PML is enabled, when the dirty bit is set, the MMU logs the guest physical address (GPA) of the accessed page to a PML buffer. Hypervisor reads the PML buffers to update the dirty page bitmap. The hypervisor needs clear the dirty bits in EPT first to ensures that writing a page will set its dirty bit. In addition, the hypervisor also needs flush the corresponding TLBs (including regular TLB and ETLB~\cite{pml-inteldoc}) to enforce the 2D page walk. The CPUs of VMs are simulated by the VCPU structure. Each VCPU has a PML buffer of 512 entries (i.e., a $4$KB page). The processor examines the PML index before logging. If the PML index is not in the range $0-511$, there is a page-modification log-full event and a VM Exit occurs~\cite{pml-inteldoc}. In particular, the hypervisor flushes the PML buffer to update the dirty page bitmap in every VM Exit~\cite{pml-flush}. In our work, the Intel PML mechanism is utilized twice: capturing the bit-settings of GPTs and optimizing dirty handling of page migration like VM live migration.

\section{Key Designs of HMM-V}
\label{design}

This section presents the key designs of HMM-V in detail. 
%HMM-V is a hypervisor-level manager for heterogeneous memory in virtualization.
%Figure~\ref{fig:arch} shows the high-level overview of HMM-V.
%HMM-V consists of four parts: PML-based page tracking, bucket sorting-based hot/cold classifier, PML-based page migration and  VM memory pool.

\subsection{PML-based VM Page Tracking}
In the page memory management, setting A/D bits in the page tables is the classic mechanism to transfer memory access information from hardware to software. 
However, traditional designs, such as the EPT scanner adopted by {\it RAMinate}, suffer from two challenges. First, the large memory footprint of the VM results in massive page table to scan every time regardless of the working set size of the applications running on the VM. Because the EPT scanner has no idea about which pages are accessed, it has to scan the entire VM mappings to examine A/D bits of pages, which results in a long scan time. Second, setting A/D bits and flushing TLB slow memory access down, especially in virtualization with costly 2D address translation. To solve the above two problems, we propose a PML-based GPT scanner and a multi-level queue design.

\subsubsection{GPT scanner based on PML.}
\label{gptscanner}

HMM-V utilizes PML to track accessed GPT pages, rather than scanning the entire EPT of the VM. HMM-V customizes a scan kernel module in the VM without modifying the gOS native code. The scan module first scans the GPTs of the processes of the VM and stores the last-level guest page table (LL-GPT) pointers to a buffer, call the LL-GPTP buffer. We use hashing to filter duplicated pointers due to page sharing. With a {\it hypercall}, the address of the LL-GPTP buffer is passed to the hypervisor. 

At initialization, the GPT scanner first clears the A/D bits of each guest page table entry (PTE) by accessing the last-level GPT pages. Then, the GPT scanner clears the corresponding dirty bits in the EPT and flushes TLBs for those GPT pages. This is to ensure that modification of GPTs can be logged by PML, as described in §\ref{subsec:pml}.
%loop
The GPT scanner in the hypervisor scans the GPT periodically and the interval between two scans is called a monitoring window. We conduct a sensitivity study of the monitoring window size (MWS) in §\ref{sub:ex-pagetrack}. During a monitoring window, the GPT scanner captures the PML log from the PML buffers in real time, and extracts the GPT page addresses from the log.
At the end of the monitoring window, the GPT scanner scans those GPT pages logged by PML, records A/D bits that are set and clears A/D bits with the multi-level queue filter discussed next. Just like initialization, the GPT scanner clears the corresponding dirty bits in EPT and flush TLBs for those GPT pages. After several rounds of monitoring (8 in our evaluation), the GPT scanner records the number of reads and writes per VM page. The pages that are not accessed do not incur A/D bits setting. Thus, the GPT scanner only scans the active GPT pages, which greatly reduces the amount of page table scanning. 

\subsubsection{Multi-level queue.}
\label{queue}
In a monitoring window, HMM-V filters each page through a multi-level queue which determines whether clear the page's A/D bits. The multi-level queue handles time events in the unit of monitoring window. At the beginning, all pages are in the level 0. The A/D bits of the pages in level 0, need to be cleared right now. For level $x~(x>0)$, we define the {\it do-not-disturb} (DND) time is {\it $2^{x-1}$}, which that means the A/D bits of pages in level $x$ will not be cleared until {\it $2^{x-1}$} intervals later.
The pages in their DND time are treated as continuous visits. When a page's DND time ends, its A/D bits should be cleared in this interval. In the next interval, if this page is set A/D bits again, it suggests the page is active (hot). Thus, we upgrade its level to make its DND time longer. Otherwise, it means that the page has not been accessed continuously, so we degrade its level to reduce its DND time. If the level of a page drops to 0, we clear the A/D bits for the page.

The multi-level queue is based on memory access locality~\cite{aet,PACE}. The hot pages have a long DND time, and the queue filters out their A/D bits setting, and estimates access counting correctly, while the cold pages in low-level queues have a short DND time, which avoids being mistakenly filtered out and missing counts.

\subsubsection{Clock switch.}
\label{sub:switch}
%间歇式监控
%基于队列思想的反馈开关等等
HMM-V turns on VM page tracking periodically through a clock switch. Experimentally, we turn on every minute for regular page VMs and 2 minutes for VMs enabling THP. In addition, HMM-V also collects the history of page migration with counters for VMs to determine whether tracking is enabled next period. When the memory migrated is under a threshold (256 MB in our evaluation), HMM-V increases the counter up to an upper limit (3 in our evaluation). When the clock switch is on, the counter decreases by one. The page tracking of a VM is enabled only if its counter is 0. The feedback mechanism avoids unhelpful page tracking and migration when most pages have been placed correctly.

%------------------------------------------------------------------------------------
\subsection{Hot Page or Cold Page?}
\label{sec:hotorcold}
The PML-based VM page tracking counts read and write frequency for each VM page. Then, HMM-V quantifies the ``temperature'' of a page (i.e., {\it page-degree}) and bucket-sorts VM pages based on {\it page-degree} in order to determine the VM hot set. 

\subsubsection{{\it Page-degree}.}
\label{subsec:pagedegree}
As shown in Table~\ref{tab:dramvs.nvm}, there is a significant gap between read and write on bandwidth and latency. To better distinguish between hot pages and cold pages, we weight read and write differently. We use micro-benchmarks (including random read/write and sequential read/write) to estimate the weights of read and write. We use the ratio of read-slowdown to write-slowdown as the ratio of the read and write weights. Specifically, read (or write) slowdown is calculated by the difference of execution time of the read (or write) benchmarks running on DRAM and NVM. We also repeat the experiment for programs with different working set sizes, and the same result shows that the ratio of read and write weights is about $1:3$. Further, we define {\it page-degree} to be $read\_wight*read\_count+write\_wight*write\_count$. Thus, we quantify page access. 

\subsubsection{Degree-sorting: distinguishing hot/cold pages}
\label{sub:sorting}

HMM-V sorts all pages of the VM based on the {\it page-degree}. According to the formula above, the {\it page-degree} is a integer with finite range. Thus, HMM-V applies the bucket sorting algorithm~\cite{bucketsort}. The sorting is done by scanning all the pages at once ($O(n)$). The hot set is built according to the DRAM capacity of the VM. Then, by using PML-based page migration, HMM-V exchanges NVM pages in the hot set and the same number of the coldest pages of DRAM.  

%------------------------------------------------------------------------------------
\subsection{Page Migration Based on Intel PML}
\label{sec:pm}

In HMM-V, DRAM and NVM serve as two different NUMA nodes for VMs. In fact, Linux kernel provides a system API~\cite{migrate-code} for migrating pages between two NUMA nodes. However, it adopts serial migration thus has low speed. {\it Nimble}'s parallel and concurrent THP migration significantly increases throughput. However, {\it Nimble} unmaps pages before copying data to ensure consistency, which results in long pauses of page accesses.  Then, {\it HeMem} write-protects pages to be migrated, and reading pages will be not interrupted during the migration. However, write protection is not virtualization friendly because of triggering heavy VMTraps. 

In contrast, HMM-V migrates pages in parallel (4 threads in our evaluation) with little interrupting page accesses, as well as, there are few VMTraps because of leveraging PML. As described in §\ref{subsec:pml}, the PML mechanism can efficiently track dirty pages for VM live-migration. Similarly, we utilize PML to track dirty pages for page migration between of DRAM and NVM. 

%Our design
After migration preparation, HMM-V first applies for new pages from the target NUMA node. HMM-V pre-applies some free pages in parallel when tracking VM pages, which speed up allocation of the new pages, especially for THP. Second, HMM-V cleans the D bits in EPT for the old pages to be migrated and flushes the corresponding TLB entries to ensure that the PML can track them. 
Third, HMM-V copies the data from the old pages to the new pages. Then, HMM-V unmaps the old pages and update the dirty page bitmap by checking the PML logs. For non-dirty pages, HMM-V sets up their new mappings directly. For dirty pages, HMM-V recopies them. By now, the mappings of dirty pages have been removed, so recopying does not incur dirty pages again. After that, HMM-V sets up their new mappings. In addition, we carefully control the number of pages migrated in parallel at the same time to mitigate the impact on VM applications due to bandwidth usage.

In particular, we actively fill the mappings of new pages in EPT after migration. To build the new page mappings, the system has to remove old mappings in the page table of QEMU and the EPT of the VM. However, remapping new pages with system API only restores the mappings in the page table of QEMU, but the mappings in EPT will be only established in the hypervisor when the VM accesses the pages again, which leads to expensive VMTraps. Thus, 
HMM-V actively fills the mappings of new pages in EPT, like the EPT fault handler.

\subsection{Memory Pool for DRAM Balancing}
\label{sec:pool}

Memory overcommit in a virtualized heterogeneous memory system suffers from new challenges comparing to traditional single-DRAM system. The large-capacity NVM guarantees the total VM memory capacity, but fast DRAM is scarce. Efficient use of DRAM is the key to maintain overall performance. Static memory allocation may incur unbalancing and low utilization. HMM-V achieves DRAM balancing across multiple VMs with a memory pool that holds free DRAM and NVM. A VM can obtain DRAM from the pool or release DRAM to the pool.

The first step in pooling management is to determine the hot set for each VM.
HMM-V obtains the {\it page-degree} distribution of each VM by bucket sorting. HMM-V defines the hot set as the top $80\%$ of pages whose page-degree is above a base threshold (We use an empirical value of $3$ in evaluation). Sorting ensures that the selected pages are always the hottest, and the base threshold ensures that a VM with many inactive pages will not be identified to have a big hot set. HMM-V sets the upper and lower DRAM limits for each VM, which can be flexibly configured by the administrator. In our experiment, we define $75\%$ and $125\%$ of the initial DRAM size as the lower and upper limits, respectively. HMM-V specifies that the hot page size ({\it hss}) is limited between the upper and lower limits.
%The second step is determining the DRAM capacity of a VM based on its hot set.
If the {\it hss} is bigger than the DRAM size, HMM-V increases the DRAM size to the {\it hss}, when the DRAM of pool is enough. If not enough, HMM-V takes as much as the pool can supply and increases the DRAM size accordingly. By migrating pages of VM NVM to the pool DRAM and remapping the new DRAM pages to the VM, HMM-V achieves adding the VM DRAM capacity. Because the pages of pool are free, the migration is one-way. If the {\it hss} of the VM is less than current DRAM size, HMM-V reduces the DRAM size to the {\it hss} and releases DRAM to the memory pool. Similarly, HMM-V migrates VM DRAM pages to the pool NVM and remaps the NVM pages to the VM.

\section{HMM-V Implementation and Evaluation}

In this section, we present the implementation first (§\ref{implement}) and evaluate HMM-V's key components (§\ref{sub:ex-pagetrack}, §\ref{sub:ex-pagemigration}, §\ref{sub:ex-ablationstudy}). Then we perform the system-level evaluation of HMM-V and compare HMM-V's performance with that of NUMA balancing ({\it NUMA\_B}) and Intel {\it MM}~\cite{OptaneDC}, as well as the {\it Nimble Page Management (Nimble)}~\cite{yan2019nimble}.  

{\it NUMA\_B} indicates that NUMA balancing is enabled in VMs without other management. In particular, we configure {\it NUMA\_B} allocating DRAM memory preferentially by using
libnuma~\cite{libnuma,gaud2014large}. {\it MM} must configure all DRAM as the DRAM cache. Thus, to be fair, we compared HMM-V with {\it MM} in scenarios where the same DRAM usage can be guaranteed. %such as multi-VM co-running.
{\it Nimble} is a heterogeneous memory management design for the THP system and we implement it in guest OS for comparison, though it is designed for non-virtualization.
{\it HeMem}, a user-level library, is difficult to put into use in the full virtualization. Therefore, we try to compare components of {\it HeMem}, including hot/cold page policies based on fixed threshold mechanism and write protection-based page migration.
%nimble占了一些便宜，在guest中不会发生vmtraps

\subsection{Implementation}
\label{implement}
HMM-V is a pure software system and we implement HMM-V in 6500 lines of C code. HMM-V adapts the QEMU/KVM~\cite{qemu} as virtualization architecture. We implement HMM-V in the KVM module of Linux kernel (5.4.142) except that an initialization scan module of page tracking is implemented in the VM as a kernel module. 
In the Intel Optance DC APP direct mode, we configure the PMem as a new NUMA node. We pre-allocate memory for each VM with two NUMA nodes.  Initially, memory is allocated from the DRAM and NVM node of the physical machine, respectively.

\subsection{Experimental Setup}
\label{sub:ex-setup}
We run our evaluation on a dual-socket Intel Cascade Lake-SP system running at 2.2 GHz with 24 cores/48 threads per socket. Each socket has 32 GB of DDR4 DRAM and 256 GB of Intel Optane DC PMem (NVM). All VMs are pinned to a single socket. Both the physical machine and the VMs run Ubuntu 18.04 with Linux Kernel 5.4.142.  
In addition, for the experiments in regular page and huge page systems, the VMs are configured with 8-cores and 16-cores, respectively.

\subsection{Page Tracking Efficiency}
\label{sub:ex-pagetrack}

Page tracking is the first step in HMM-V. The accuracy of page tracking significantly affects the distinction between hot and cold pages and the overhead of page tracking can affect the VM performance.

\subsubsection{Page table scanning efficiency.}
\begin{figure}[ht]
\begin{center}
\includegraphics[width=0.48\textwidth]{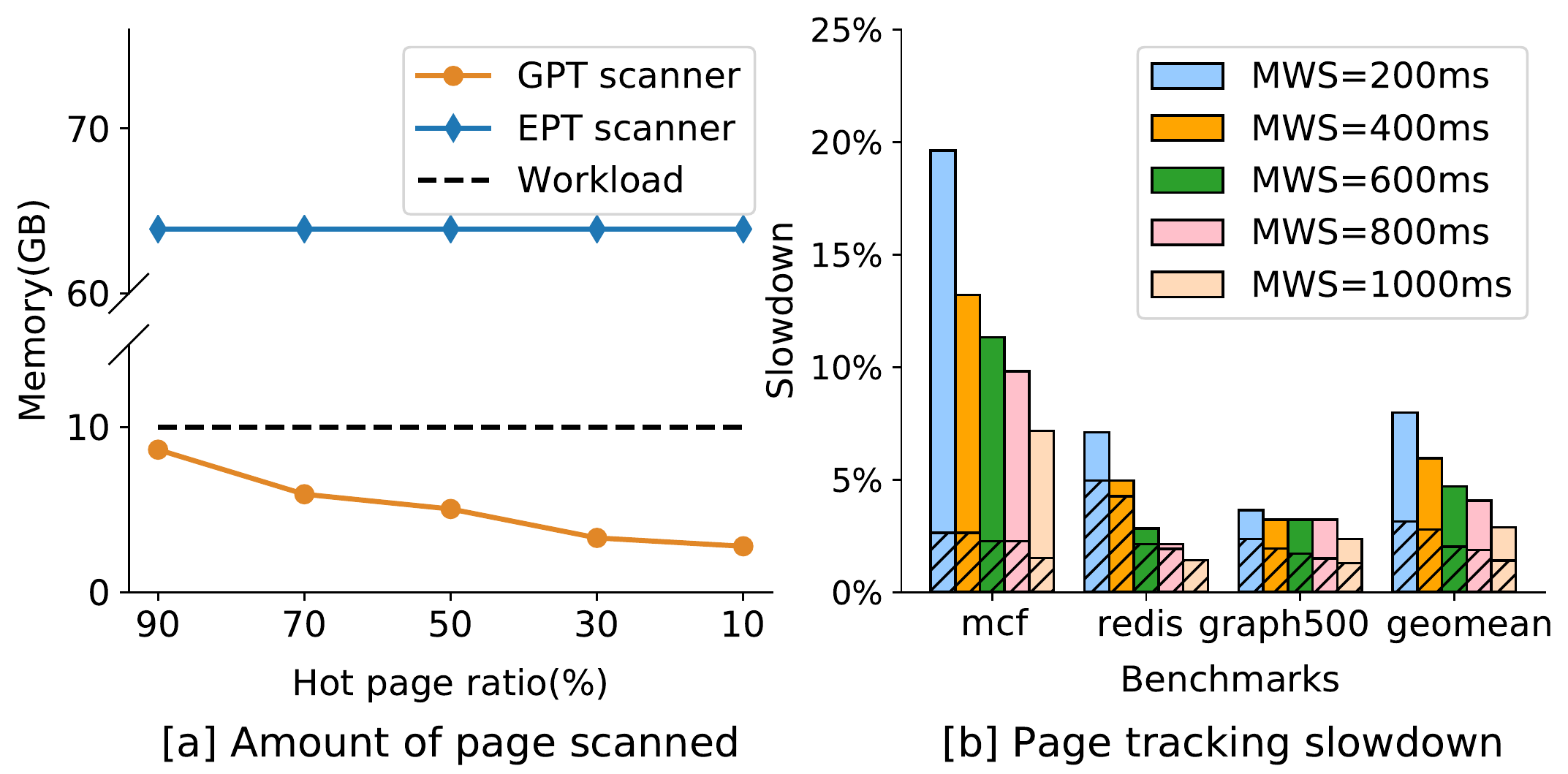}
\caption{[a] Comparison of page scanned (dotted line represents the RSS of the workload); [b] Comparison of page tracking slowdown (solid line columns represent EPT scanner and shadow columns represent GPT scanner). }   
\label{fig:e-page_tracking}
\end{center}
\vspace{-0.5cm}
\end{figure}

To evaluate the page table scanning efficiency of PML-based GPT scanner, we execute a \texttt{random access} workload on a VM with 64 GB of memory. The workload has 10 GB of resident set size (RSS, i.e., memory footprint). After warming-up, the VM establishes 64 GB memory mappings. Figure~\ref{fig:e-page_tracking} [a] shows the result (GPT scanner vs. EPT scanner). The X-axis represents $90\%$ access touching $X\%$  (i.e., hot page ratio) of pages and the Y-axis shows the amount of VM memory scanned. 

The results show that the GPT scanner is affected by the locality of memory accesses. The better the locality is, the fewer GPTs are involved. The GPT scanner uses PML to accurately capture GPTs accessed recently, which greatly reduces the number of pages to be scanned. In contrast, because the EPT scanner is not aware which pages are accessed, it has to scan the whole EPT (with 64 GB of mappings) regardless of the hot page ratio. It will cause long time of page tracking when the VM has a huge mapping.

%-------------------------------------------------------------------------------------

\subsubsection{Page tracking overhead.}
\label{subsub:page-tracking-overhead}

We choose \texttt{429.mcf} (with 3.2 GB RSS) of SPEC CPU2006~\cite{spec06}, \texttt{graph500} (with 19 GB RSS)~\cite{graph500} and \texttt{redis}~\cite{redis} (with 32 GB RSS) as the workloads to test the slowdown caused by page tracking. 
\texttt{Graph500} iterate 8 times with BFS and SSSP. Both \texttt{graph500} and \texttt{429.mcf} use execution time as performance metric.  We test \texttt{redis} with YCSB~\cite{ycsb} and each key-value size is 4 KB. {\it READ} and {\it UPDATE} account for $50\%$ respectively and the workload follows the {\it hotspot} distribution, where $80\%$ of access operations occur on $20\%$ of data. \texttt{Redis} uses the throughput as performance metric. We choose the EPT scanner adopted by {\it RAMinate} as a comparison (§\ref{sec:relatedwork}). In this experiment, both the GPT scanner and the EPT scanner are enabled all the time.

Frequent page table scanning results in a significant address translation overhead because of TLB flush and page table A/D bits setting. As shown in Figure~\ref{fig:e-page_tracking} [b], the EPT scanner causes up to a $20\%$ slowdown for \texttt{429.mcf}, but the GPT scanner with multi-queue causes only a $2.6\%$ slowdown.
We increase the monitoring window size (MWS) to 1000 ms (adopted by {\it RAMinate}), and the slowdown of \texttt{429.mcf} with the EPT scanner is still up to $8\%$. In contrast, the slowdown of the GPT scanner is no more than $2\%$. For applications, like \texttt{429.mcf}, with memory intensive, the GPT scanner achieves a lower slowdown by utilizing the multi-level queue to filter large amount of hot page tracking. \texttt{Redis} accesses memory sparsely with low frequency and the GPT scanner still has performance advantages than the EPT scanner. By geometric mean, our design can reduce the page tracking slowdown by more than $50\%$ over the EPT scanner. In particular, when $MWS \ge 600ms$, the slowdown is less than $2\%$. Thus, we choose 600 ms as MWS for subsequent experiments.

%pml buffer full
In addition, we also test the additional VMTraps caused by using the PML. Each VCPU has a separate PML buffer with 512 entries. When the buffer is full, an additional VMTrap, {\it PML BUFFER FULL}, is triggered. We calculate the proportion of VMTraps generated due to {\it PML BUFFER FULL} event during the execution of \texttt{random access} workload with working set from 4 GB to 16 GB. The larger the working set of the workload, the more VMTraps are generated due to {\it PML BUFFER FULL}. The results show that no more than $0.34\%$ of additional VMTraps are generated. By leveraging PML, we capture dirty GPT pages rather than all dirty pages. The hypervisor actively flushes the PML buffer every time when any type of VMTraps is triggered.

\subsubsection{Multi-level queue.}
\label{subsub:ex-queue}
The locality-based multi-level queue estimates the page access frequency. We choose multiple benchmarks to test whether the multi-level queue approach can keep the memory access pattern. Due to space limitations, we only present the result of \texttt{redis}. The RSS is 10 GB and \texttt{redis} has a fixed hot set (2 GB), which makes it easy to observe. We monitor 100 cycles and the maximum queue level is configured as 7.

\begin{figure}[h!]
	\begin{center}
		\includegraphics[width=0.48\textwidth]{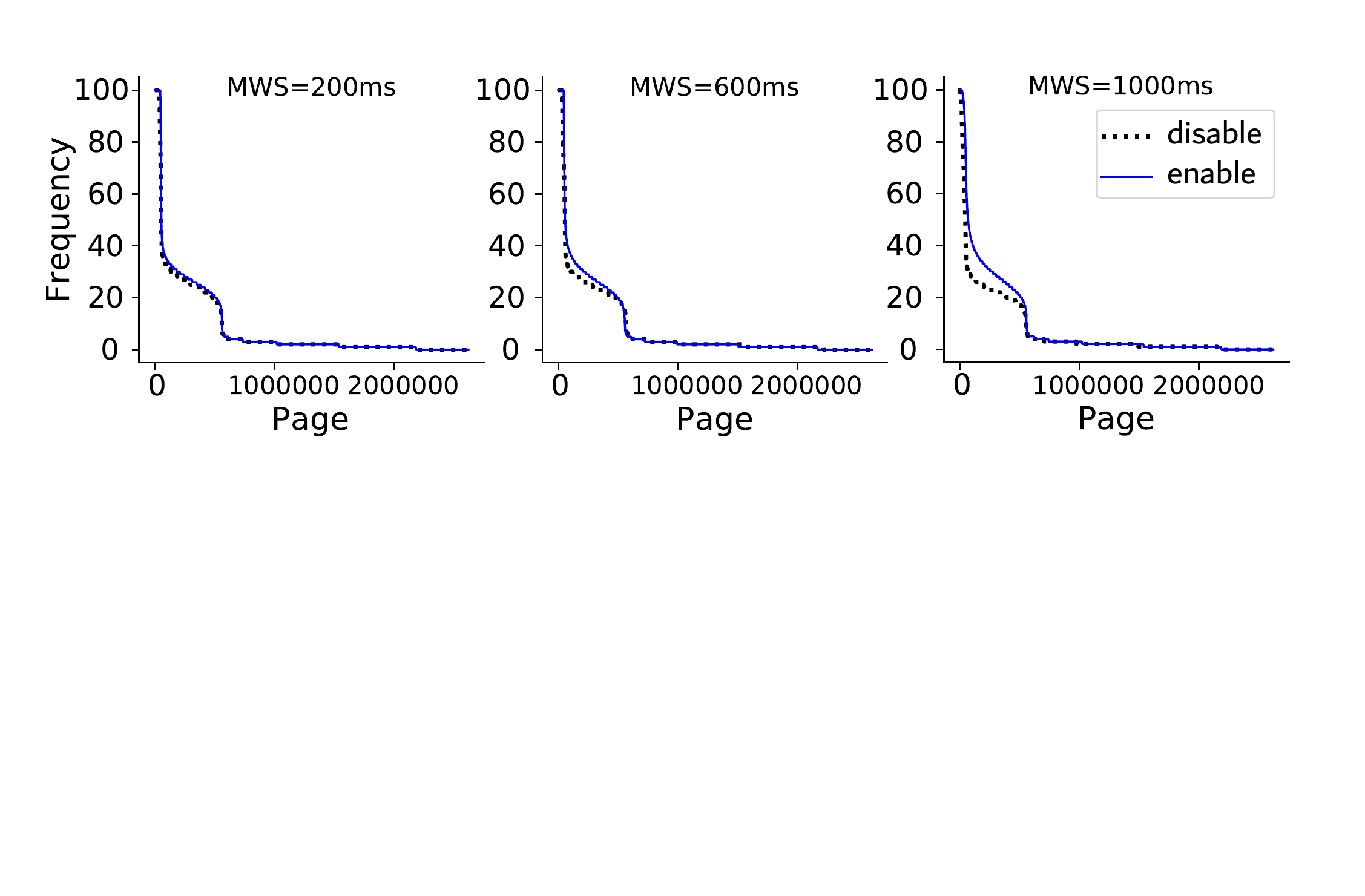}
		\caption{Access frequency of tracked pages with/without multi-level queue.} 
		\label{fig:e-queue}
	\end{center}
	\vspace{-0.8cm}
\end{figure}

Figure~\ref{fig:e-queue} shows distribution of sorted page access frequency and the monitor window size (MWS) changes from 200 ms to 1000 ms. In all the tests, the PML-based GPT scanner can capture the hot set ($20\%$) accurately. The solid and dashed lines have the same shape, indicating that the memory access pattern is described correctly when enabling multi-level queue. We observe that the frequency of hot pages is higher in the distribution when enabling multi-level queue. Hot pages enter a higher-level queue and their A/D bits are not be cleared in the next $2^{i-1}$ monitor windows, so that they are projected to accessed $2^{i-1}$ times although the actual frequency might be less. Cold pages are not affected because they stay in the low level queues and prediction error is much smaller. 
The multi-level queue makes it more efficient to distinguish between hot and cold pages. 

In addition, when the MWS is increased, the frequency increase of hot pages caused by multi-level queue becomes more significant.
%4kb value; throughput is 3000 opt/s; hot set is 2GB (i.e., 2^19 pages)  
According to the throughput measured by YCSB, \texttt{redis} accesses memory pages with very low frequency. Thus, with a small MWS, a page cannot be captured in multiple continuous monitor windows, which makes the level upgrade difficult. Therefore, when MWS is $200ms$, there is no difference whether multi-level queue is enabled or not. Conversely, with bigger MWS, accesses to a same page are more likely to be captured in consecutive windows and the level upgrade will be faster. Combined with the results of §\ref{subsub:page-tracking-overhead}, we can conclude that multi-level queue can effectively reduce the overhead caused by page table scanning while maintaining the application memory access pattern.

\subsection{Page Migration Efficiency}
\label{sub:ex-pagemigration}

In this section, we compare PML-based page migration (PM\_PML) to raw Linux page migration (PM\_Linux), and write protection-based page migration (PM\_WP) adopted by {\it HeMem}. We should focus not only on migration speed but also on the slowdown caused by migration. We configure an VM with 20 GB DRAM and 20 GB NVM. We choose benchmarks with uniform random memory access, including \texttt{random read} and \texttt{random write}. We run a benchmark on the NVM first and then migrate all pages of the benchmark to the DRAM.

\subsubsection{Page migration speed.}
\label{subsub:pmspeed}
Obviously, the sooner we can finish the migration, the sooner hot pages can be accessed on high-speed DRAM. %Table~\ref{tab:pmspeed} presents the experimental results of migration times. 
We vary the RSS of the workload from 2 GB to 16 GB and observe that the migration speed (throughput) is stable. The results show that benefiting from parallel migration, PM\_PML and PM\_WP have basically the same speed and both are twice as fast as PM\_Linux. PM\_PML and PM\_WP migrate a write-workload is slower than a read-workload because of handling dirty pages.

\iffalse
\begin{table}[htbp]
\vspace{-0.4cm}
  \centering
  %\tabcolsep=0.1cm
  \renewcommand\arraystretch{0.8}
  \caption{Page migration time(s).}
    \begin{tabular}{ccccccc}
    \toprule
        RSS  & \multicolumn{2}{c}{PM\_Linux} & \multicolumn{2}{c}{PM\_WP} & \multicolumn{2}{c}{PM\_PML} \\
    (GB)  & read  & write & read  & write & read  & write \\
    \midrule
    2     & 5     & 6     & 2     & 3     & 2     & 3 \\
    4     & 10    & 11    & 5     & 5     & 4     & 5 \\
    8     & 19    & 21    & 8     & 9     & 9     & 10 \\
    16    & 40    & 42    & 18    & 23    & 18    & 23 \\
    \bottomrule
    \end{tabular}%
  \label{tab:pmspeed}%
 \vspace{-0.3cm}
\end{table}%
\fi

\subsubsection{Slowdown caused by page migration.}
We use the execution time of the workload as a metric to measure the slowdown of the workloads caused by page migration. 
%Figure~\ref{fig:e-pm} 
Table~\ref{tab:pm-read} and Table~\ref{tab:pm-write} show the workload slowdowns (compared to DRAM performance) after migration by PM\_PML, PM\_WP, PM\_Linux. We also present the slowdowns (compared to DRAM performance) of running on NVM without migration.

%\iffalse
\begin{table}[htbp]
  \centering
  \tabcolsep=0.1cm
  \caption{Slowdown caused by migration (random-read).}
    \begin{tabular}{ccccc}
    \toprule
    RSS (GB) & NVM   & PM\_Linux & PM\_WP & PM\_PML \\
    \midrule
    2     & 62.8\% & 6.9\% & 0.7\% & 0.1\% \\
    4     & 60.6\% & 7.0\% & 2.8\% & 1.5\% \\
    8     & 72.1\% & 7.6\% & 3.1\% & 1.7\% \\
    16    & 78.0\% & 8.0\% & 3.3\% & 2.2\% \\
    \bottomrule
    \end{tabular}%
  \label{tab:pm-read}%
\end{table}%
%\fi

For \texttt{random read}, both PM\_PML and PM\_WP have better performance than PM\_Linux. 
Both PM\_PML and PM\_WP migrate pages faster and do not need to unmap pages before copying page data, which reduces the pause time for VM page access during page migration. Also, when running \texttt{random read}, there are few dirty pages to handle. PM\_PML causes lower slowdowns than PM\_WP because PM\_PML actively fills mappings in the EPT, which reduces the cost of VMTraps to handle EPT page faults. The results of VMTraps caused by page migration are shown in Table~\ref{tab:vmtraps_pm}. For \texttt{random read}, both PM\_Linux and PM\_WP generate VMTraps that are roughly equivalent to the number of migrated pages. PM\_PML, in contrast, incurs few VMTraps. 

%\iffalse
\begin{table}[htbp]
  \centering
  \caption{Slowdown caused by migration (random-write).}
   \tabcolsep=0.1cm
    \begin{tabular}{ccccc}
    \toprule
    RSS (GB) & NVM   & PM\_Linux & PM\_WP & PM\_PML \\
    \midrule
    2     & 188.2\% & 4.8\% & 0.7\% & 0.3\% \\
    4     & 190.6\% & 5.0\% & 3.6\% & 1.8\% \\
    8     & 189.8\% & 5.4\% & 5.7\% & 2.5\% \\
    16    & 183.5\% & 10.0\% & 8.4\% & 3.4\% \\
    \bottomrule
    \end{tabular}%
  \label{tab:pm-write}%
\end{table}%
%\fi

For \texttt{random write}, both PM\_Linux and PM\_WP cause much a higher slowdown than PM\_PML. For example, when migrating the 16 GB workload, PM\_Linux and PML\_WP cause 3.2x and 2.5x slowdowns over PM\_PML, respectively. Compared to PM\_Linux, PM\_PML finishes migration faster due to parallel processing. As shown in Table~\ref{tab:vmtraps_pm}, compared to migrating \texttt{random read}, PM\_WP doubles the number of VMTraps when migrating write workloads of the same RSS. PM\_WP triggers a large number of VMTraps because of write-protection exceptions during page migration. PM\_PML, in contrast, eliminates this overhead by leveraging PML.

\iffalse
\begin{figure}[t]
\vspace{-0.3cm}
\begin{center}
\includegraphics[width=0.48\textwidth]{pic/e-pm.pdf}
\caption{Slowdown caused by page migration. }   
\label{fig:e-pm}
\end{center}
\vspace{-0.5cm}
\end{figure}
\fi

\begin{table}[h!]
\vspace{-0.3cm}
  \centering
   \renewcommand\arraystretch{0.9}
  \caption{VMTraps caused by page migration (thousand)}
    \begin{tabular}{ccccccc}
    \toprule
    RSS  & \multicolumn{2}{c}{PM\_Linux} & \multicolumn{2}{c}{PM\_WP} & \multicolumn{2}{c}{PM\_PML} \\
    (GB)  & read  & write & read  & write & read  & write \\
    \midrule
    2     & 523   & 524   & 526   & 1030  & <1    & <1 \\
    4     & 1039  & 1049  & 1051  & 1988  & <1    & <1 \\
    8     & 2097  & 2080  & 2102  & 3927  & 1     & 1  \\
    16    & 4188  & 4196  & 4203  & 7721  & 2     & 2  \\
    \bottomrule
    \end{tabular}%
  \label{tab:vmtraps_pm}%
 \vspace{-0.3cm}
\end{table}%
\subsection{Ablation Study}
\label{sub:ex-ablationstudy}

In this section, we evaluate the performance improvements generated by the three main components (i.e., PML-based page tracking, bucket sort-based hot/cold classifier and PML-based page migration) of HMM-V. 
We replace HMM-V components with EPT scanner, a fixed threshold classifier, and PM\_Linux for comparison. In particular, the fixed threshold classifier compares a page's access count with a fixed threshold to directly determine whether the page is a hot page. Following the method of {\it HeMem}~\cite{raybuck2021hemem}, we select the value of fixed threshold by profiling a \texttt{random access} benchmark.
In addition, we run benchmarks in a pure DRAM VM for comparison. We configure the VM with 8 GB DRAM and 32 GB NVM except the pure DRAM system.

\begin{figure}[ht]
\begin{center}
\includegraphics[width=0.48\textwidth]{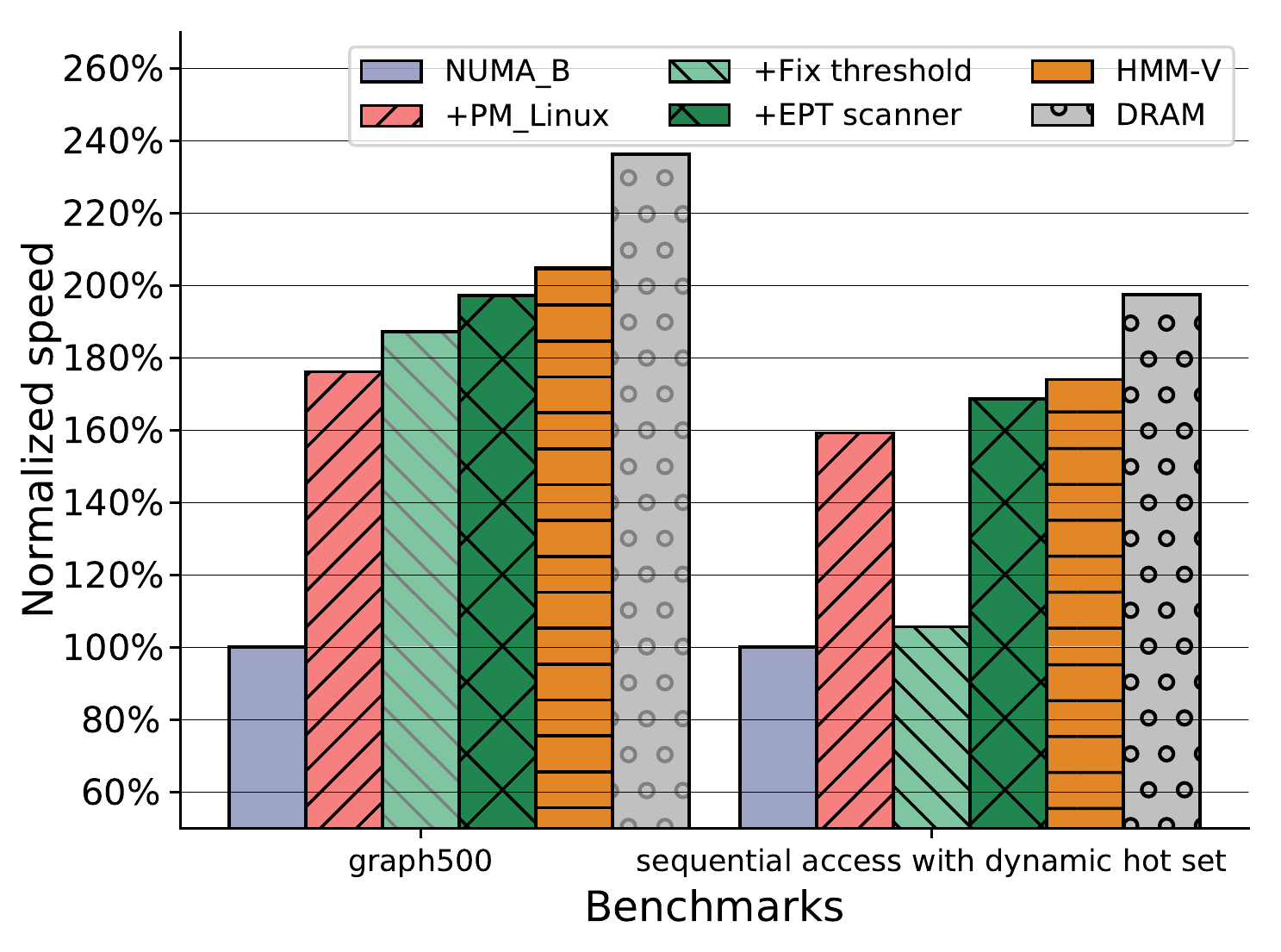}
\caption{Benchmark execution performance (the performance of {\it NUMA\_B} is normalized as 1).}   
\label{fig:ablationstduy}
\end{center}
\vspace{-0.5cm}
\end{figure}

Figure~\ref{fig:ablationstduy} shows the relative performance speedup of these six configurations over {\it NUMA\_B}.  We choose \texttt{graph500} (with 19 GB RSS) and \texttt{sequential access} (with 36 GB RSS) as workloads. The \texttt{sequential access} consists of $50\%$ read and $50\%$ write and its hot set size is fixed (7.2 GB), but the hot data distribution changes periodically.
For \texttt{graph500} and \texttt{sequential access}, HMM-V outperforms {\it +EPT scanner} by $3.9\%$ and $3.2\%$, respectively. This indicates that the PML-based GPT scanner with multi-level queue has less page tracking overhead than the EPT scanner.
For \texttt{graph500}, HMM-V outperforms {\it +fixed threshold} by $9.4\%$ and for \texttt{sequential access}, the performance speedup is about $65\%$. Distinguishing hot/cold pages based on sorting adopted by HMM-V is more adaptable than the fixed threshold classifier.
HMM-V provides $16.3\%$ (\texttt{graph500}) and $9.3\%$ (\texttt{sequential access}) higher performance than {\it +PM\_Linux}.  With PML-based page migration, HMM-V benefits from high page migration speed and efficient dirty page handling mechanism that minimizes the pause time for VM page access during page migration. Because of more intensive memory write access, for HMM-V, \texttt{sequential access} costs more to handle dirty page than \texttt{graph500}.
In addition, HMM-V outperforms {\it NUMA\_B} $105\%$ and $74\%$ for \texttt{graph500} and \texttt{sequential access}, respectively. HMM-V can provide more than $87\%$ DRAM VM system performance with only $20\%$ DRAM both in two tests.

\subsection{Fixed/Dynamic Hot Set}
\label{sub:ex-hotset}
 HMM-V aims to place the hot set into DRAM accurately. We verify HMM-V's adaptability to the VM hot set in this section. We use a \texttt{random access} micro-benchmark and set $90\%$ of operations to access hot objects while the remaining $10\%$ of operations uniformly access the entire memory footprint. The experimental VM is configured with 8 GB DRAM and 16 GB NVM. 

\subsubsection{Fixed hot set.} 
\begin{figure}[ht]
\begin{center}
\includegraphics[width=0.48\textwidth]{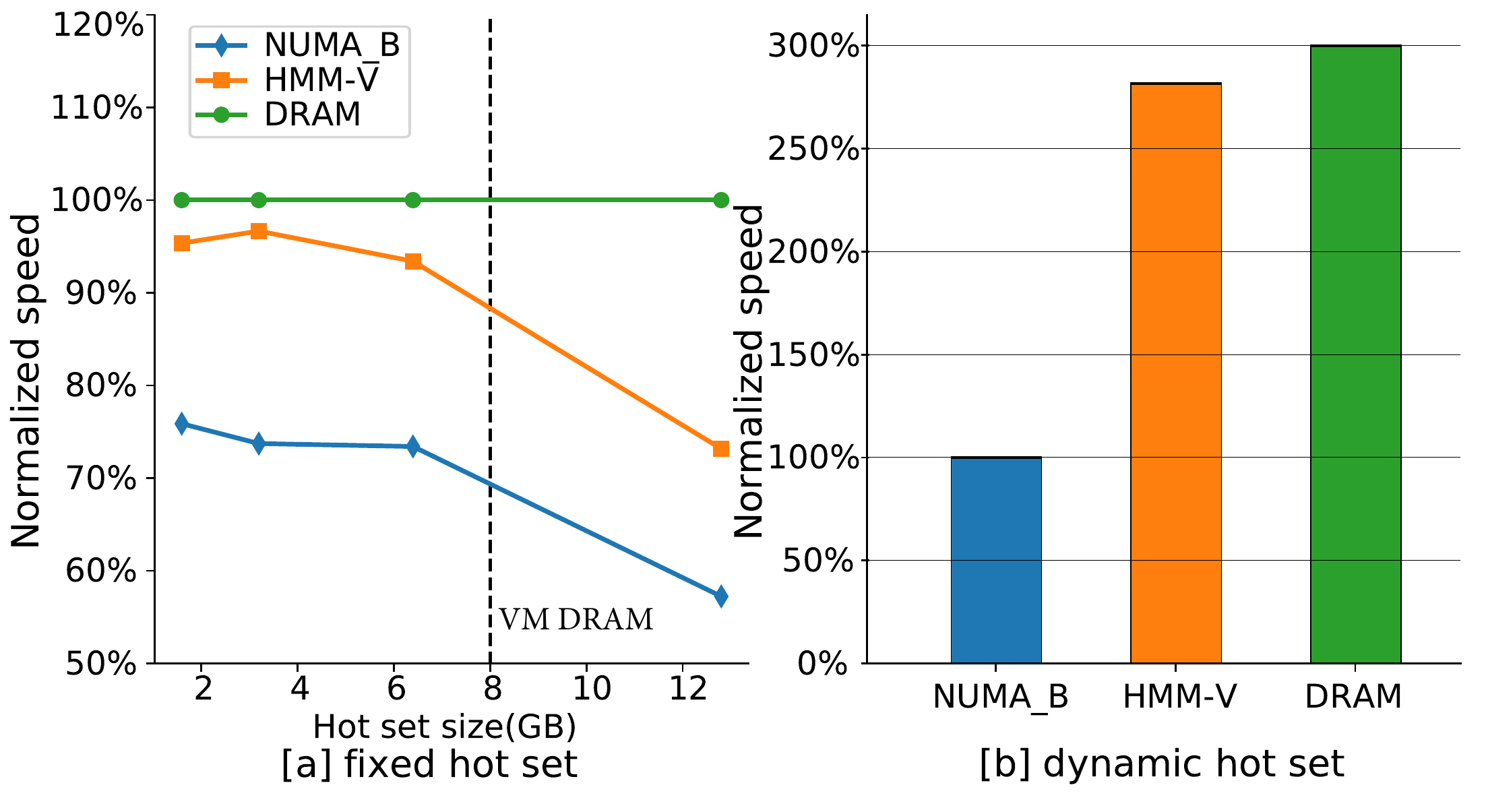}
\caption{Micro-benchmark performance.} 
\label{fig:hotset}
\end{center}
\vspace{-0.5cm}
\end{figure}

First, we configure the micro-benchmarks with fixed hot sets of different size. As shown in Figure~\ref{fig:hotset} [a], we vary the hot set ratio of the workload (with 16 GB RSS) from $10\%$ to $80\%$. 
When the hot set can fit in the DRAM, HMM-V can keep more than $93\%$ performance compared to the pure DRAM system, which indicates that HMM-V can identify the hot set and migrate hot pages into DRAM memory rapidly. The performance gap comes from approximately $10\%$ of accesses must go to slow NVM. On the contrary, the performance of {\it NUMA\_B} suffers an average loss more than $25\%$ than the DRAM system. When the hot set exceeds the DRAM capacity, the VM performance with HMM-V and {\it NUMA\_B} converges. But HMM-V still outperforms {\it NUMA\_B} by $28\%$. When there are no cold pages in DRAM to exchange hot pages in NVM, HMM-V stops page migration.

\subsubsection{Dynamic hot set.} 
We configure the \texttt{random access} with a dynamic hot set. The benchmark has four stages and in each stage, the hot data is distributed in a different 4 GB virtual address space randomly. As shown in Figure~\ref{fig:hotset} [b], HMM-V achieves $94\%$ performance of the pure DRAM system, but {\it NUMA\_B} suffers from a $1.5\times$ slowdown over the pure DRAM system. HMM-V can accurately track hot pages and migrate them into DRAM in time even when the hot set changes dynamically.

%\fi

\subsection{Sensitivity to DRAM Memory Size}
\label{sub:ex-dramsize}
\begin{figure}[h!]
\begin{center}
\includegraphics[width=0.48\textwidth]{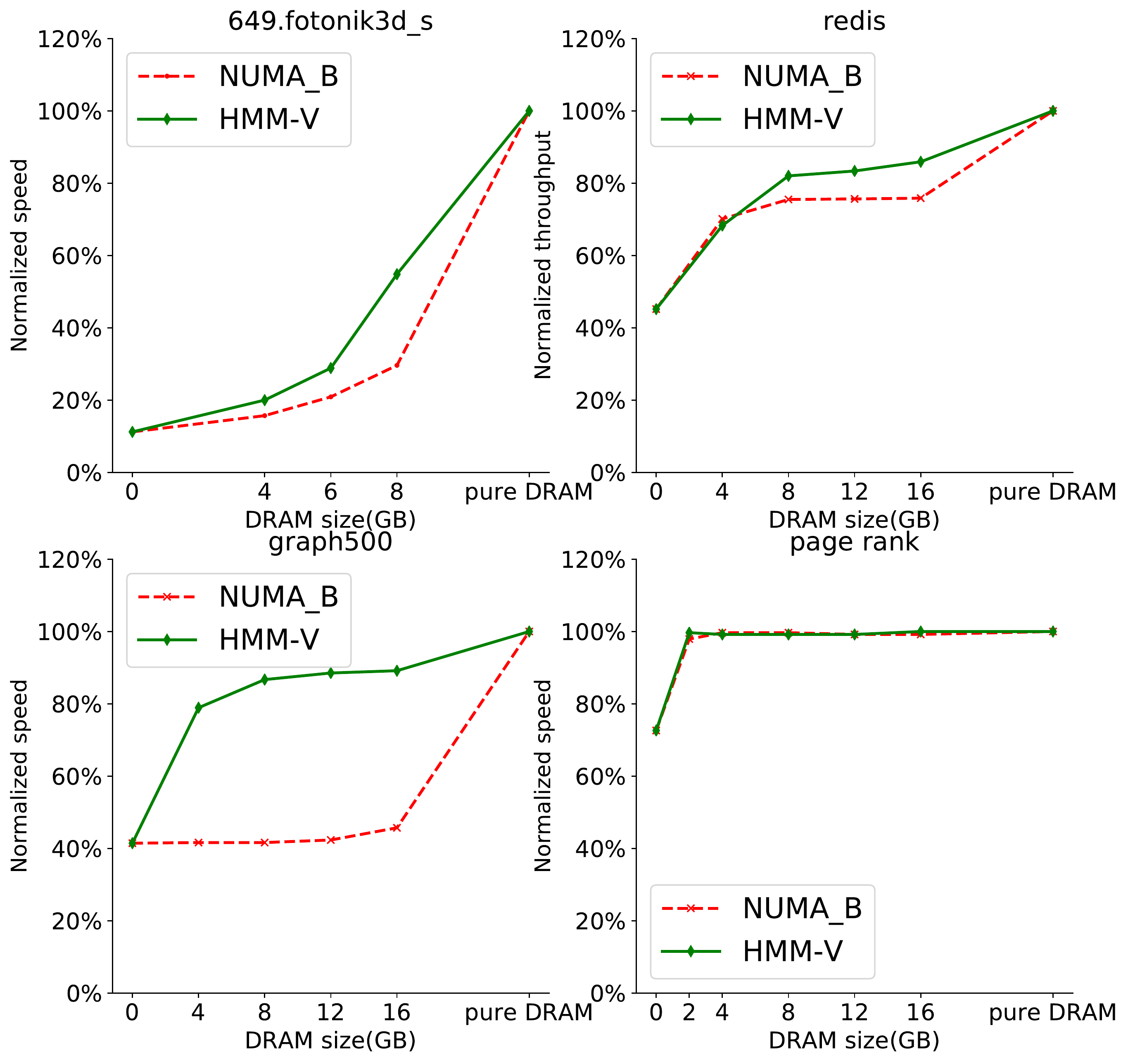}
\caption{VM performance sensitivity to DRAM size.} 
\label{fig:dramsize}
\end{center}
\vspace{-0.5cm}
\end{figure}

Effective use of DRAM is the key to ensure the performance of heterogeneous memory system. In practice, VMs may be configured with different DRAM memory capacities for cost-performance ratio. In this section, we vary the VM DRAM size and show the adaptability of HMM-V. We choose the NUMA\_B as a baseline. We select four workloads: \texttt{649.fotonik3d\_s} of SPEC CPU 2017~\cite{spec17}, \texttt{graph500}, \texttt{redis} and \texttt{page rank} of GAPS~\cite{gap}. As shown in Figure~\ref{fig:dramsize}, different types of applications have different sensitivity to DRAM size.

%649.fotonik3d\_s
As a parallel CPU and memory test benchmark, \texttt{649.fotonik3d\_s} intensively accesses the entire working set partition. Both HMM-V and {\it NUMA\_B} incur a high slowdown when the memory footprint (10 GB) exceeds the DRAM size. This indicates that \texttt{649.fotonik3d\_s} has a very urgent and huge demand for fast memory (DRAM).
%graph500
For \texttt{graph500}, when the DRAM size varies from 4 GB to 16 GB, HMM-V reaches $79\%$ to $90\%$ performance of that of the pure DRAM VM. HMM-V outperforms {\it NUMA\_B} $95.6\%$ on average. This benefits from HMM-V's ability to determine hot set accurately and migrate hot pages from NVM to DRAM quickly.
%redis
We configure \texttt{redis} with 32 GB RSS and $20\%$ hot data. As an in-memory database with a large memory footprint, \texttt{redis} has a sparse memory access pattern. When the DRAM size is smaller than hot set size, both HMM-V and NUMA\_B provide about $70\%$ DRAM performance due to frequent NVM access. Once DRAM could fit hot page set, HMM-V achieves better performance (more than $10\%$) than NUMA\_B. 
%page rank
\texttt{Page rank} mainly computes the contribution of each node by sparsely walking the graph node arrays in sequence. Thus, it has poor locality. The memory footprint of \texttt{page rank} is configured as 19 GB. HMM-V achieves $100\%$ DRAM performance when $DRAM\geq 2 GB$. By analyzing the access distribution of VM pages with HMM-V, we observe that most memory accesses are located in the top of address space, a small amount of memory ($\leq 1 GB$). {\it NUMA\_B} uses fast memory preferentially, so hot set fit perfectly in DRAM and it also achieves $100\%$ of DRAM performance.

\subsection{Transparent Huge Page (THP) Support}
\label{sub:ex-thp}

This section evaluates HMM-V performance when the VM enables THP mechanism. We compare the performance of HMM-V with that of {\it NUMA\_B}, {\it MM} and {\it Nimble}. {\it MM} must use 32GB of DRAM as DRAM cache. To be fair, for HMM-V, {\it NUMA\_B} and {\it Nimble}, we configure VM with 32 GB DRAM. Benchmarks used include page rank algorithm (\texttt{PR}) and betweenness centrality (\texttt{BC}) algorithm of GAPS~\cite{gap}, \texttt{graph500} and \texttt{redis}. The parameters and memory footprints of benchmarks are shown in Table~\ref{tab:thpbenchs}.  We provide two sets of parameters for each benchmark. For example, we configure \texttt{page rank} with $2^{27}$ vertices (fits in DRAM) and $2^{29}$ vertices (exceeds DRAM), respectively, called \texttt{PR\_S} and \texttt{PR\_L}.

\begin{table}[htbp]
  \centering
  \vspace{-0.3cm}
  \caption{Parameters and RSSes of benchmarks.}
  \label{tab:thpbenchs}%
  \setlength\tabcolsep{2.5pt}
  \begin{threeparttable}
    \begin{tabular}{ccccccccccc}
    \toprule
     & \multicolumn{2}{c}{$PR^{+}$} & \multicolumn{2}{c}{$BC^{+}$} & \multicolumn{2}{c}{graph500}  & \multicolumn{2}{c}{redis} \\
    \midrule   
    \specialrule{0em}{1pt}{1pt}
    $Parameter^{*}$ & $2^{27}$ & $2^{29}$ & $2^{27}$  & $2^{29}$   & $2^{25}$  & $2^{27}$   & 8M  & 25M \\
    {RSS (GB)} & 18    & 70    & 20    & 76    & 19    & 76     & 32    & 100 \\
    \bottomrule
    \end{tabular}%
  \begin{tablenotes}
        \footnotesize
        \item[*] For \texttt{PR}, \texttt{BC}, \texttt{graph500}, the parameter represents the number of vertices; For \texttt{redis}, it represents the number of K-V pairs.
        \item[+] For \texttt{PR} and \texttt{BC}, we present the RSS during the stable iteration period, and building the graph requires doubling the RSS. 
  \end{tablenotes}
    \end{threeparttable}
\vspace{-0.3cm}
\end{table}%

\begin{figure}[t]
\begin{center}
\includegraphics[width=0.48\textwidth]{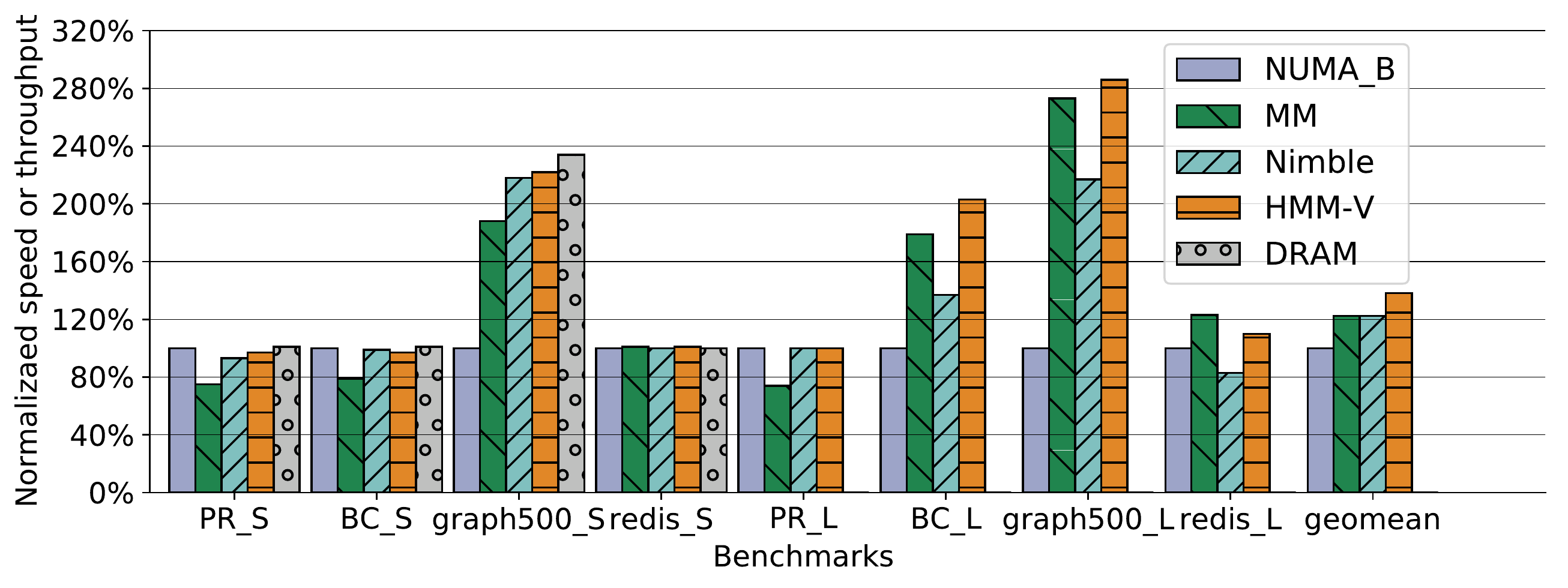}
\caption{Benchmark performance when VM enables THP (the performance of {\it NUMA\_B} is normalized as 1).} 
\label{fig:singlevmthp}
\end{center}
%\vspace{-0.5cm}
\end{figure}

Figure~\ref{fig:singlevmthp} presents the performance comparison. When memory footprint can fit in DRAM, HMM-V remains the workload's all memory in DRAM, achieving performance close to the pure DRAM system. HMM-V outperforms {\it MM} by $14.7\%$ on average. With a direct-mapped DRAM cache, MM can suffer conflict misses incurring costly NVM accesses and the frequent NVM write back is also expensive, especially for write-intensive programs like \texttt{graph500} and \texttt{BC}. We configure {\it NUMA\_B} allocating DRAM memory preferentially by using libnuma. However, \texttt{graph500\_S} tries to allocate memory from two NUMA nodes equally for CPU balancing. As a result, {\it NUMA\_B} provides only half the performance of the DRAM system. In addition, like HMM-V, {\it Nimble} migrates all  workload pages and achieve performance close to the pure DRAM system.

%exceed the DRAM
When the memory footprint exceeds the DRAM capacity, the workloads behave differently. 
%page rank
For \texttt{PR\_L}, HMM-V and {\it Nimble} provide the same of performance with {\it NUMA\_B}. This result matches the analysis of sensitivity to DRAM size in §\ref{sub:ex-dramsize}. However, {\it MM} suffers from a $26\%$ slowdown due to heavy conflict cache misses.
%bc graph500
For \texttt{BC\_L}, {\it Nimble} outperforms {\it NUMA\_B} by $30\%$ and HMM-V achieves $48.4\%$ higher performance than {\it Nimble}. HMM-V achieves page tracking with low overhead, and benefiting migration feedback mechanism (§\ref{sub:switch}), HMM-V can suspend unnecessary page scans based on history. By contrast, we observe that for program with large working set and intensive memory access pattern, the performance of {\it Nimble} is limited due to the high frequency of page scanning without optimization. Compared with {\it MM}, HMM-V provides $13\%$ higher performance for \texttt{BC\_L}. In addition, \texttt{graph500\_L} has basically the same results as \texttt{BC\_L}.

%redis
\texttt{Redis\_L} is configured with 20 GB hot set, which can fit in DRAM. HMM-V outperforms {\it NUMA\_B} by more than $10\%$. HMM-V identifies hot pages and migrate into DRAM. {\it MM} outperforms HMM-V by about $10\%$. The hot keys of redis distribute randomly in entire address space and the redis access memory sparsely. Therefore, when memory footprint exceeds DRAM capacity, {\it MM}, a fine-grained (cache line) hardware management, is able to exploit more locality than page-level management. In addition, {\it Nimble}'s performance is even $19\%$ lower than the baseline ({\it NUMA\_B}).
Nimble adopts active/inactive LRU lists of the Linux memory reclamation mechanism~\cite{yan2019nimble}. {\it Two page states ("active" and "inactive') are not sufficient to describe all possible access patterns}~\cite{bovet2005understanding}. For example, although a page of DRAM is "inactive" for the most of time, one recent access makes it "active", thus denying migrating the page into NVM, even if it is not going to be accessed for a long time. Redis has this kind of pattern.

%silo
%Silo is an in-memory transactional database and we run TPC-C model with 16 threads, which simulate customers making orders from warehouses. The access pattern is random and there is little read and write reuse. When RSS can fit in DRAM, HMM-V is able to achieve up to $6\%$ higher throughput than MM because HMM-V remains all RSS in DRAM. Conversely, when the RSS of silo exceeds DRAM capacity, HMM-V and NUMA\_B have similar performance and MM outperforms them about $19\%$. In test of redis and silo, Compared to baseline (NUMA\_B), {\it Nimble} provides a negative effect, because the LRU-based hot and cold page replacement is not suitable for sparse and random access.

\subsection{Multi-VM Co-Running}
\label{sub:ex-multi-vm}

This section evaluates HMM-V when multiple VMs co-run, including regular page and THP. We configure HMM-V as two modes: HMM-V island mode and HMM-V pool mode. In island mode, each VM has a fixed DRAM capacity and HMM-V independently manages pages for each VM.  In pool mode, HMM-V enables memory pooling (§\ref{sec:pool}) and DRAM memory is balance on demand.

\subsubsection{VMs with Regular Page.}
\begin{figure}[t]
\begin{center}
\includegraphics[width=0.48\textwidth]{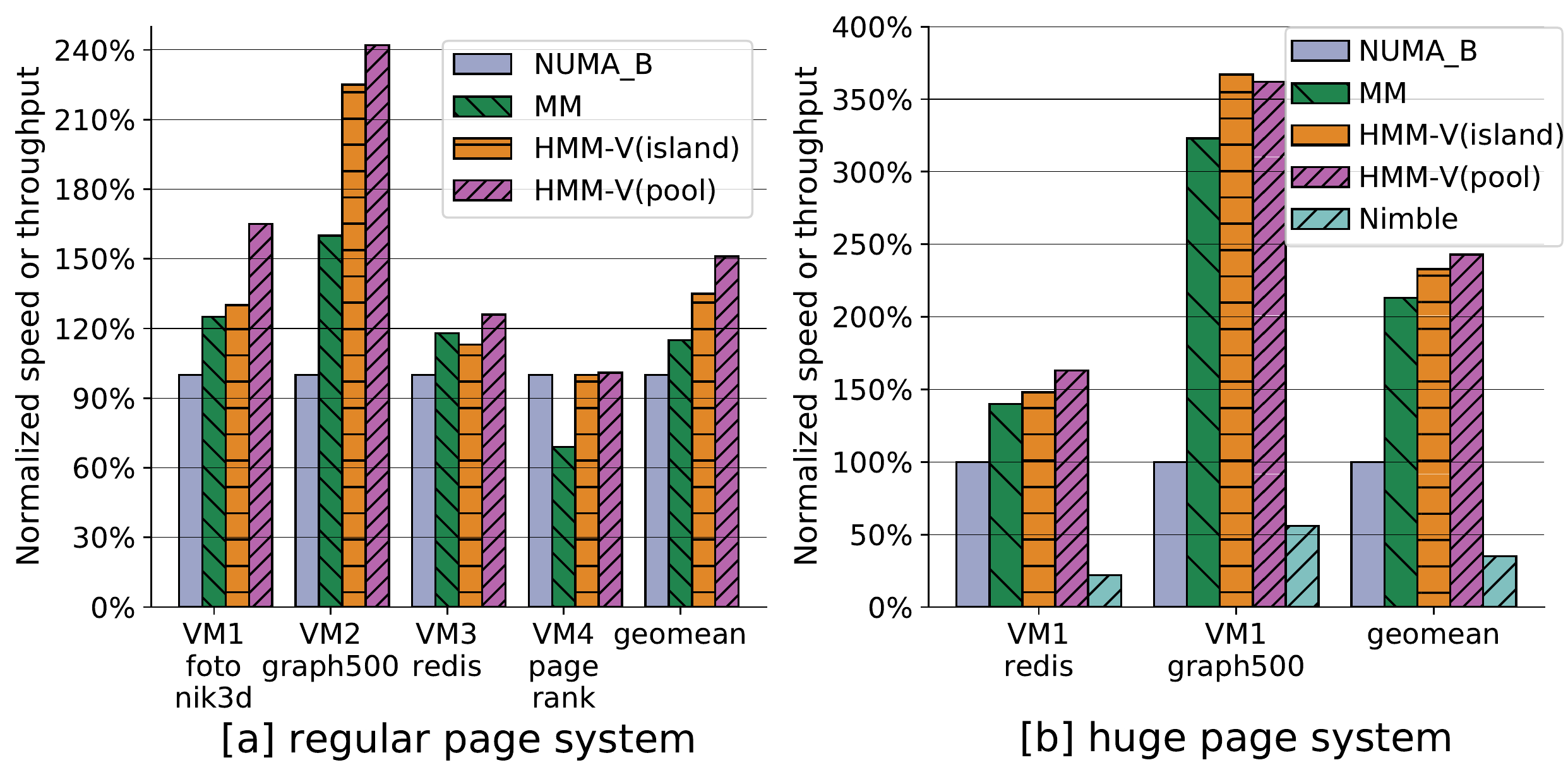}
\caption{Benchmark performance in multi-VMs.} 
\label{fig:multi_vms}
\end{center}
\vspace{-0.5cm}
\end{figure}

We configure four VMs and each VM has 40 GB memory. The initial DRAM size of each VM is 8 GB. For pool mode, and we set $75\%$ of the initial DRAM size as the default (6 GB). This is the lower bound of DRAM size. The benchmarks are configured the same as~§\ref{sub:ex-dramsize}. Figure~\ref{fig:multi_vms}[a] shows the result. 

Using {\it NUMA\_B} as baseline, by geometric mean, {\it MM} provides a $15\%$ performance improvement. However, {\it MM} suffers from a $30\%$ slowdown for the VM4 (\texttt{page rank}), which matches the result of §\ref{sub:ex-thp}. Moreover, the performance degradation is exacerbated because multi-VM co-running causes heavier DRAM cache pollution. 
Overall, HMM-V island mode outperforms {\it NUMA\_B} and MM by $35\%$ and $17\%$ (geomean), respectively. When multiple VMs co-run, HMM-V still achieves distinguishing hot/cold pages and relocating pages effectively. And HMM-V effectively avoids the interference of DRAM access of multiple VMs. 

The overall performance of HMM-V pool mode is $31\%$ and $12\%$ higher than {\it MM} and HMM-V island mode, respectively. In particular, HMM-V pool mode outperforms {\it MM} and HMM-V island mode on every VM. 
We observe that with management of pool mode, VM2, VM3 and VM4 release 1.5 GB, 1.5 GB and 2 GB to the pool respectively, while VM1 gets 1.5 GB DRAM memory from the pool. Like the analysis in §\ref{sub:ex-dramsize}, \texttt{649.fotonik3d\_s}, with 10 GB RSS, makes intensive access to the most address space, which requires more DRAM memory. \texttt{Page rank} running in VM4 requires little DRAM, so the DRAM is down to the default value (6 GB). The hot set of \texttt{redis} is 6.4 GB, so it frees nearly 1.5 GB of DRAM. VM2 (\texttt{graph500}) also progressively releases 1.5 GB DRAM. Thus the VMs actually consume a total of 28.5 GB DRAM. % Pool is a logical concept. The pool DRAM memory is actually still in a {\it freelist} of the host OS and can be allocated to other applications. 
In particular, the pool mode achieves higher performance but with less DRAM for VM2 and VM3 than the island mode.  For VM2 and VM3, the remaining DRAM is sufficient to cover their hot sets. Without sufficient DRAM, under HMM-V island mode, VM1 suffers frequent page migration, which affects the performance of other VMs by consuming memory bandwidth. In summary, with pooling management, HMM-V achieves higher performance by judiciously allocating precious DRAM memory.

\subsubsection{VMs with enabling THP.}
\label{subex:multivmwiththp}
%single VM
%\texttt{graph500}  3.88% \texttt{redis} 6.52%    bc 8.86%
%co-run
%\texttt{graph500} 14.87% \texttt{redis} 16.89%
%                \texttt{redis}  14.92%  bc 13.22% 

We evaluate HMM-V when VMs enable THP and the performance comparison is shown in Figure~\ref{fig:multi_vms}[b]. Each VM has 128 GB memory including 16 GB DRAM. For pool mode, we also set $75\%$ of the initial DRAM size as the default (i.e., $12GB$). VM1 runs \texttt{redis} with 50 GB RSS (including $20\%$ hot data) and VM2 executes \texttt{graph500} with 76 GB RSS. 

When the VMs enable THP, HMM-V still achieves higher performance than MM.
HMM-V (island) and HMM-V (pool) outperform MM $9.3\%$ and $13.9\%$ (geomean), respectively. We observe that when running \texttt{redis} in the single VM, MM performs better than HMM-V island mode (see §\ref{sub:ex-thp}). But when two VMs co-run, HMM-V (island) outperforms MM $5.3\%$ on \texttt{redis}. Similarly, the performance gap between MM and HMM-V (island) widens for \texttt{graph500}. We measure the DRAM cache load miss ratio for MM by detecting two hardware events\footnote{Intel PMU does not provide similar events for store instruction}: $MEM\_LOAD\_RETIRED.LOCAL\_PMM$ (PMM\_C) and $MEM\_LOAD\_L3\_MISS\_RETIRED.LOCAL\_DRAM$ (DRAM\_C)~\cite{pmu}.
The former counts retired load instructions with local PMem as the data source and the data request missed the DRAM cache in MM. The latter counts retired load instructions with data serviced from local DRAM. Thus, the DRAM cache load miss ratio is $PMM\_C/(PMM\_C+DRAM\_C)$.
We compare miss ratio when the workload runs in single VM and double VMs. For \texttt{redis}, the miss ratio increases from $6.52\%$ to $16.89\%$ and for \texttt{graph500}, it increases from $3.88\%$ to $14.87\%$. The result indicates that when multiple VMs co-run, {\it MM} suffers heavy (conflict) cache misses. 

HMM-V (island) and HMM-V (pool) outperform {\it Nimble} $5.6$x and $5.8$x (geomean), respectively. In particular, {\it Nimble} provides a lower performance on \texttt{graph500} compared to {\it NUMA\_B}, which is the opposite of single-VM result (§\ref{sub:ex-thp}). The DRAM size of the single-VM experiment is double that of the multi-VM experiment. The larger the DRAM size, the higher the wrong page placement tolerance. {\it Nimble}'s LRU-based policies cannot accurately identify hot set when DRAM size is relatively limited.
However, both HMM-V island and pool modes achieve more than $5\times$ performance over {\it Nimble}, which suggests that HMM-V still works well under the condition of tight DRAM size, due to our page monitoring and bucket sort-based hot/cold page classifier.

\section{Conclusion}
This paper proposes HMM-V, a novel heterogeneous memory (DRAM+NVM) management system for virtualization. HMM-V holds hot pages in fast DRAM and cold pages in NVM through page tracking, classification, and migration. In particular, HMM-V optimizes performance by utilizing hardware-assisted virtualization. In addition, HMM-V adopts memory pooling management to balance DRAM between multiple VMs for higher utilization and performance.
We have implemented HMM-V in a real system that supports Intel Optane DC PMem, and conducted a systematic evaluation. Experimental results show that HMM-V outperforms MM, {\it Nimble} and NUMA balancing when multi-VM co-running. 

\bibliographystyle{plain}
\bibliography{main}

\end{document}